\documentclass[a4paper, 12pt, aps,reprint,longbibliography,nofootinbib, pra]{revtex4-1}

\usepackage{physics}
\usepackage{graphicx}
\usepackage{dcolumn}
\usepackage{bm}
\usepackage{bbold}

\begin{document}


\title{Local Probability Conservation in Discrete Time Quantum Walks}

\author{Samuel T. Mister}
 \email{sm15883@my.bristol.ac.uk}
\author{Benjamin J. Arayathel}%
 \email{ben.arayathel@gmail.com}
\author{Anthony J. Short}
 \email{tony.short@bristol.ac.uk}
\affiliation{H. H. Wills Physics Laboratory, University of Bristol, Tyndall Avenue, Bristol, BS8 1TL, United Kingdom}

\begin{abstract}
We show that probability is locally conserved in discrete time quantum walks, corresponding to a particle evolving in discrete space and time. In particular, for a spatial structure represented by an arbitrary directed graph, and any unitary evolution of a particle which respects that locality structure, we can define probability currents which also respect the locality structure and which yield the correct final probability distribution. 
\end{abstract}

\maketitle


\section{\label{sec:Intro}Introduction}

\noindent For a  particle evolving via the Schrodinger equation in continuous space and time, it is well known that any changes in its probability density can be explained by local probability currents. This result has recently been extended to discrete space and continuous time \cite{schumacher_westmoreland_new_qiao_2016}. In this paper we will demonstrate that this is also the case for discrete space and time, hence ensuring local conservation of probability for discrete time quantum walks \cite{kempe, kendon, Ash07}. 

In continuous space and time the local conservation of probability for a single particle is expressed by the continuity equation
\begin{equation}
    \frac{\partial \rho}{\partial t} + \nabla \cdotp J = 0, 
    \label{eqn:CSCT_CON}
\end{equation}
where $\rho = |\psi|^2$ is the probability density and $J$ is a  vector field describing the probability current. For a particle  governed by the non-relativistic Schr\"{o}dinger equation $i \hbar \frac{\partial \psi}{\partial t} = -\frac{ \hbar^2 }{2m}\nabla^2 \psi + V \psi$  we find that
\begin{equation}
    J = -\frac{i \hbar}{2m}[\psi^* \nabla \psi - \psi\nabla\psi^*],
    \label{eqn:CSCT}
\end{equation}
is real and satisfies equation \eqref{eqn:CSCT_CON}. From this we can conclude that probability is conserved locally in this case. A similar probability current can be defined for relativistic systems governed by  the Dirac equation \cite{dirac}. 

The same is true if we make space discrete. In this picture we represent space as a graph. Then the continuity equation representing  local conservation of probability becomes 
\begin{equation}
    \frac{dP_n(t)}{dt} + \sum_m J_{mn} = 0, 
    \label{eqn:DSCT_CON}
\end{equation}
where $P_n$ represents the probability of being at vertex $n$ and $J_{mn}$ is a matrix element representing the probability current between vertexes $n$ and $m$ (where $J_{mn}>0$ implies a net flow of probability from $n$ to $m$). To ensure locality we require that $J_{mn}=0$ whenever $n$ and $m$ are not linked by an edge in the graph, and in order to obtain meaningful results, we also require that $J_{mn}$ be real and anti-symmetric.
It has been shown that for any system undergoing Schr\"{o}dinger evolution with Hamiltonian $H$, we can take $J_{mn}$ to have the  form \cite{schumacher_westmoreland_new_qiao_2016} 
\begin{equation}
    J_{mn} = \frac{1}{i}(H_{mn}\rho_{nm}  - \rho_{mn}H_{nm}).
    \label{Schumacher_Current}
\end{equation}
Here $\rho$ represents the density operator of the particle. Given that $H_{mn}$ and $H_{nm}$ are zero  whenever $n$ and $m$ are not linked by an edge, $J$ is a local probability current which satisfies \eqref{eqn:DSCT_CON} and is real and antisymmetric. Hence again in these systems  probability is locally conserved. 

We now take this further by also making time discrete. Instead of the Schr\"{o}dinger equation, in each time step a unitary operator is applied to the state, such that  $\ket{\psi'} = U \ket{\psi}$. This  corresponds to a discrete time quantum walk. As time derivatives are not applicable in this case, the continuity equation \eqref{eqn:DSCT_CON} must be modified to refer to the change in probability $\Delta P_n = P^{'}_{n} - P_{n}$ at vertex $n$ in one time step, and the probability current $J_{mn}$ flowing between $n$ and $m$ in one time step, giving 
\begin{equation}
    \Delta P_n + \sum_m J_{mn} = 0.
    \label{eqn:DSDT}
\end{equation}
There are four main properties that the probability current $J$ should satisfy. As in the previous case  it should be real, anti-symmetric and non--zero only when $n$ and $m$ are connected by an edge in the graph. However, here an additional property to enforce locality is required - that the probability flux out of a given vertex in one time step is less than the initial probability of being at that vertex. This property can be written concisely as
\begin{equation}
    \sum_{m^*}J_{m^{*}n} \leq P_n,
    \label{eqn:Prob_condition}
\end{equation}
where $m^* = \{m:J_{mn} > 0\}$. We will use this notation for $m^*$  throughout the paper. An expression for $J$ which satisfies the first three properties has been proposed \cite{schumacher2}, but does not satisfy \eqref{eqn:Prob_condition}. In this paper we show that a valid probability current satisfying all four conditions can be found in all cases, thus ensuring local probability conservation for discrete space and time. We also extend these results to cases with internal degrees of freedom and directed graphs, for which we require that $J_{mn}>0$ only if there is a directed edge from $n$ to $m$. 

\section{Setup}

\begin{figure}
    \centering
    \includegraphics[scale = 0.5]{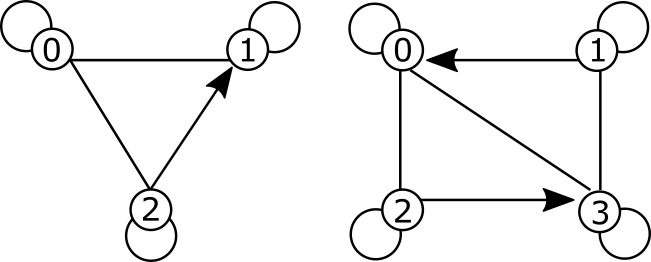}
    \caption{Examples of directed graphs that allow discrete time Quantum walks. Edges without arrows are undirected and can be traversed in either direction.}
    \label{fig:Directed_graphs}
\end{figure}

\noindent A suitable description of our discrete space is a graph, consisting of a set of vertices $V$ and a set of edges $E$. For full generality, we consider directed graphs, for which an edge is associated with a particular direction of travel. Examples of these types of graphs are shown in figure \ref{fig:Directed_graphs}. These graphs allow us to include novel space-time structures in which the particle is restricted to travel in certain directions. An edge is specified by an ordered pair of vertices $E \subseteq \{n \rightarrow m\, | n,m \in V\}$. For example, the edge $n \rightarrow m$ would allow the particle to move from $n$ to $m$. We assume that the particle is always allowed to remain at its current location, so all self loops are included in $E$ ($n \rightarrow n \in E$ for all $n$) \footnote{If we allow graphs for which some self loops are not included, then we can still prove local probability conservation using the probability flow approach given in the next section. However, the corresponding restrictions on the probability current are more complicated.}.  To restrict to the simpler case of undirected graphs, we would require that $n \rightarrow m \in E \implies m \rightarrow n \in E$.

The time evolution of a quantum particle in our discrete space-time model corresponds to a discrete time quantum walk on this graph. To define such a quantum walk, we associate an orthonormal quantum state $\ket{n}$ to each vertex (corresponding to the particle being at that point), and specify a unitary operator $U$ describing the evolution, for which the matrix elements $U_{mn} = \bra{m} U \ket{n}$ satisfy $n \rightarrow m \notin E \implies U_{mn}=0$. Hence the unitary evolution cannot move the particle between vertices which are not connected by an edge. Given an initial pure state $\ket{\psi}$, we have $P_n= |\langle n | \psi \rangle|^2$ and $P_n'= |\langle n | U | \psi \rangle|^2$ 

Note that in the case of discrete space and continuous time, it is unnecessary to consider directed graphs, because if $n \rightarrow m \notin E$ and thus $H_{mn} = 0$, the Hermitian nature of the Hamiltonian means that $H_{nm} = 0$ and therefore the directed edge in the opposite direction $m \rightarrow n$ cannot be used during the evolution. 

In the discrete time case, directed graphs can lead to interesting results, and have been previously studied in the context of discrete time quantum walks. In particular it has been shown that reversibility of the graph is a necessary and sufficient condition to define a coined Quantum walk \cite{Ash07}. An edge from $n \rightarrow m$ is reversible if there exists a path from $m$ to $n$, and a graph is reversible if every edge in it is reversible. We extend our results to coined quantum walks, and other cases in which the vertices have internal states, in section \ref{sec:internal}.

\section{Results}

\subsection{Probability flow}

\noindent In order to analyse the locality of probability flows, it is helpful to break the probability current $J_{mn}$ (which represents the net flow of probability between vertices $n$ and $m$) into the individual flows of probability along the directed edges $n \rightarrow m$ and $m \rightarrow n$. In particular, we define the flow of probability along the edge $n \rightarrow m$ as $f_{mn}$. Then 
\begin{equation}
    J_{mn} = f_{mn}-f_{nm}.
    \label{Current_Recovery}
\end{equation}
Note that the `diagonal' flow matrix element $f_{nn}$ corresponds to the amount of probability which remains at vertex $n$.

In order to give meaningful results and satisfy local probability conservation, the flow matrix elements $f_{mn}$ must satisfy the following properties:
\begin{align}
    f_{mn} &\geq 0, \label{Flow1}\\
        f_{mn} &= 0 \quad \text{if} \quad n\rightarrow m \notin E,\label{Flow2}\\
    \sum_m f_{mn} &= P_n,  \label{Flow3}\\
    \sum_n f_{mn} &= P_m'.
    \label{Flow4}
\end{align}
The first condition specifies that the probability flowing along an edge in a particular direction must be positive, the second that it must respect the locality structure of the graph. The third condition specifies that all probability initially at vertex $n$ must either flow to a neighbouring vertex or remain there during one time-step. The fourth condition requires that all probability  at vertex $n$ after one time step must either have flowed to it from a neighbouring vertex or have remained there.

We now show that these properties for $f_{mn}$ yield all the required properties of $J_{nm}$. The flow $f_{nm}$ is  a positive number hence $J_{nm}$ as defined in \eqref{Current_Recovery} is real. We also see that $J_{nm}$ is anti-symmetric, non-zero only when an edge exists between $m$ and $n$, and satisfies equations \eqref{eqn:DSDT}~and~\eqref{eqn:Prob_condition}. 
\begin{align}
   \Delta P_n + \sum_m J_{mn} &= (P_n'-P_n) +\sum_m f_{mn} - \sum_m f_{nm} \nonumber \\
    &= (P_n'-P_n) + P_n - P_n' \nonumber \\ 
    &= 0
\end{align}
\begin{align}
    \sum_{m^*}J_{m^{*}n} &= \sum_{m^*}f_{m^{*}n}-\sum_{m^*}f_{nm^{*}} \nonumber\\
    &\leq\sum_{m}f_{mn} -\sum_{m^*}f_{nm^{*}} \nonumber\\
    &\leq P_n.
\end{align}
 Below, we show that a valid $f_{nm}$ satisfying  properties \eqref{Flow1}-\eqref{Flow4} always exists, hence we can also define a valid $J_{nm}$ satisfying local probability conservation.

The converse is also true. If we can define a $J_{nm}$ which is real, antisymmetric, satisfies \eqref{eqn:DSDT} and \eqref{eqn:Prob_condition}, and for which $J_{mn} > 0$ only if $m \rightarrow n \in E$ then we can always generate flows $f_{mn}$ satisfying conditions \eqref{Flow1}-\eqref{Flow4}. This is shown in the appendix, and illustrates that  flow conditions \eqref{Flow1}-\eqref{Flow4} are equivalent to the conditions on the $J_{mn}$ given in the introduction.

\subsection{Existence of local probability flows}
\noindent To prove that we can define flow matrix elements $f_{nm}$ satisfying the conditions \eqref{Flow1} to \eqref{Flow4}, we can use a result of Aaronson \cite{aaronson_2005}. However, for completeness and clarity, here we provide a simpler proof of a similar result which is sufficient for our purposes.

\begin{figure}
    \centering
    \includegraphics[scale=0.65]{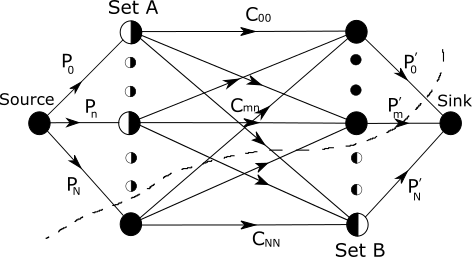}
    \caption{A diagram showing the movement of probability in the network flow picture. The line on the graph is an example of a cut. Vertices corresponding to $n \in A$ and $m \in B$ are shown half-filled.}
    \label{fig:Network}
\end{figure}

The key insight is to consider  probability as a `fluid', flowing through a network of `pipes' with different capacities from a source to a sink. This can be described by a directed graph with edges which have a maximum capacity specifying the amount of probability allowed to flow along them. Figure \ref{fig:Network} illustrates the configuration we will consider.

This network consists of 3 different groups of edges. The first and final sets of edges have capacity corresponding to initial and finial probabilities respectively. The intermediate edges represent the evolution of the state and have capacity defined in the following way \footnote{Note that the similar result in \cite{aaronson_2005} takes $C_{mn} =|U_{mn}|$. This leads  to a valid flow satisfying $f_{mn}\leq |U_{mn}|$.}
\begin{equation} 
C_{mn} = 
    \begin{cases}
    0 & \mbox{if } n\rightarrow m \notin E \\
    1 & \mbox{otherwise}
    \end{cases}
\end{equation}
If the total capacity of this network from source to sink is at least one, then for any flow configuration achieving  capacity one the flow of probability along the edges in the middle section  will give a valid $f_{mn}$. In particular, we set  $f_{mn}$ equal to the flow of probability along the intermediate edge with capacity $C_{mn}$.

Following a similar approach to \cite{aaronson_2005}, we will show that the maximum flow allowed by the network is no less than one unit of probability by making use of the max-flow, min-cut theorem \cite{Dantzig}. This states that the value of the minimum cut in the network is equal to the maximum flow of the network. A cut is a set of edges which if removed from the network disconnects the source from the sink, and its value is the total capacity of those edges. 

Let us first write down the value of a general cut. Let $A$ be the set of $n$ such that the edge $Source \rightarrow n$ is not in the cut and let $B$ be the set of $m$ such that the edge $m \rightarrow Sink$ is not in the cut. Then to disconnect the source from the sink the cut must contain all the edges $n \rightarrow m$ such that $n \in A$ and $m \in B$. Therefore the value of the cut can be written as
\begin{align}
    \sum_{n\notin A} P_n + \sum_{m \notin B} P_{m}' &+ \sum_{n\in A,m\in B}C_{mn} \nonumber \\
    = (1 - \sum_{n\in A}P_n) &+ (1 - \sum_{m \in B}P_m') + \sum_{n\in A,m\in B}C_{mn} \nonumber
\end{align}
For our claim to hold the above expression must be greater than or equal to one. To show this we prove the following inequality
\begin{equation}
    \sum_{n\in A}P_n + \sum_{m\in B} P_{m}' \leq 1 + \sum_{n\in A, m\in B}C_{mn}.  
    \label{eqn:INEQUALITY}
\end{equation}
Firstly, we consider the case in which at least one of the $C_{mn}$ elements in the cut is non zero and secondly we consider the case where all of the $C_{mn}$ elements in the cut are zero. 

 In the first case, the right hand side of \eqref{eqn:INEQUALITY} is at least two. As any partial sum over elements of a probability distribution is at most one, the sum of the two terms on the left is at most two, and the inequality is satisfied. 
 
In the second case, the right hand side of \eqref{eqn:INEQUALITY} is equal to one. In this case, it is helpful to express the left hand side of \eqref{eqn:INEQUALITY}  in terms of projection operators as
\begin{align}
    \sum_{n\in A}P_n + \sum_{m\in B} &P_{m}' =   \sum_{n \in A}|\bra{n}\ket{\psi}|^2 + \sum_{m \in B} |\bra{m}U\ket{\psi}|^2 \nonumber\\
    &= \bra{\psi}\Big( \sum_{n \in A}\ket{n}\bra{n} + \sum_{m \in B} U^\dagger\ket{m}\bra{m}U \Big) \ket{\psi}\nonumber\\
    & = \bra{\psi}\Pi_A + \Pi_B\ket{\psi}
    \label{projection}
\end{align}
where $\ket{\psi}$ is the initial state, 
\begin{align}
     \quad \Pi_A &= \sum_{n\in A} \ket{n}\bra{n} \quad \text{and } \quad \Pi_B = \sum_{m\in B} U^\dagger\ket{m}\bra{m}U.
\end{align}
$\Pi_A$ and $\Pi_B$ are projectors onto the spaces spanned by $\ket{n}$ such that $n \in A$ and $U^\dagger\ket{m}$ such that  $m \in B$ respectively. In order to show that \eqref{projection} is at most one, it suffices to show that $\Pi_A + \Pi_B$ is a projection operator.
\begin{align}
    (\Pi_A + \Pi_B)^2 &= \Pi_A^2 + \Pi_B^2 + \Pi_A\Pi_B + \Pi_B\Pi_A \nonumber\\
    \Pi_A\Pi_B &=  \sum_{n\in A,m\in B} \ket{n}\bra{n}U^\dagger\ket{m}\bra{m}U = 0 \nonumber\\  
    \Pi_B\Pi_A &=  \sum_{n\in A,m\in B} U^\dagger\ket{m}\bra{m}U\ket{n}\bra{n} = 0\nonumber\\
    \implies (\Pi_A + \Pi_B)^2 &= \Pi_A + \Pi_B
\end{align}
The cross terms go to zero as by assumption $C_{mn} = 0$ which implies $U_{mn} = 0 \implies U_{mn}^* = 0 $. It is also clear that $(\Pi_A + \Pi_B)^\dagger = \Pi_A + \Pi_B$. Hence $\Pi_A + \Pi_B$ is a projection operator and $\bra{\psi}\Pi_A + \Pi_B\ket{\psi} \leq 1$.\\

This shows that all cuts in the network shown in figure \ref{fig:Network} have value greater than or equal to one. This then implies that the minimum cut in the network has value greater than or equal to one. Then by applying the Max-flow, Min-cut theorem we can conclude that the maximum flow allowed in the network is greater than or equal to one. As we only require one unit of probability to flow through the network at each time step this is sufficient to show that there exists a valid probability flow for every discrete time Quantum walk. Hence probability is locally conserved for quantum evolutions in discrete space time. 

\subsection{Constructing solutions}

\noindent Although the above proof ensures the existence of a valid probability flow satisfying local probability conservation, it does not give a method of constructing such a flow. However, this can be achieved efficiently for cases with a finite number of vertices via linear programming. 

If $N$ is the number of vertices in $V$, we can think of the flow matrix elements $f_{nm}$ as forming an $N^2$ dimensional real vector $\mathbf{f}$. The constraints \eqref{Flow1}-\eqref{Flow4} then correspond to a positivity constraint on each component of $\mathbf{f}$, and a number of linear equalities satisfied by the components. These can be expressed in the form 
\begin{align}
\mathbf{f} &\geq 0, \\
\mathbf{A}. \mathbf{f} &= \mathbf{b}, 
\end{align} 
where $\mathbf{A}$ and $\mathbf{b}$ are a matrix and vector expressing the linear equalities \eqref{Flow2} - \eqref{Flow4}. 
Given such constraints, a linear program can find a vector $\mathbf{f^*}$ which satisfies the constraints and maximizes the value of some linear objective function $c=\mathbf{v}.\mathbf{f}$. In this case, as we are only interested in finding a feasible assignment $\mathbf{f}$, it does not really matter what we choose as our objective function, but one natural choice would be to maximize the amount of probability which remains stationary (i.e. taking $c=\sum_n f_{nn}$). This would prevent probability from flowing in both directions between two vertices. 

Various techniques exist to solve linear programming problems, including the simplex method \cite{simplex}, or Karmarkar's  algorithm \cite{Karmarkar}. The latter approach is efficient in the computational complexity sense, requiring a time which is polynomial in $N$.

\subsection{Systems with Internal Degrees of Freedom} \label{sec:internal}

\noindent Quantum systems with internal degrees of freedom are commonly used  in the context of coined quantum walks. In particular, we could consider a particle which carries an internal degree of freedom, such as a spin, in addition to its location. Alternatively we could consider cases in which each spatial location has its own distinct set of internal states. 

In both of these cases we can denote an orthonormal basis of quantum states by $\ket{n,k}$ where $n\in V$ gives the spatial location and $k \in \mathcal{S}_n$ gives the internal degree of freedom. In such cases, we can apply the results obtained earlier, and thus prove local probability conservation, by mapping the system to one with no internal degrees of freedom. In this mapping, a vertex with $M$ internal degrees of freedom can be replaced with a set of $M$ vertices that are all connected to each other.

\begin{figure}[h]
\includegraphics[scale = 0.5]{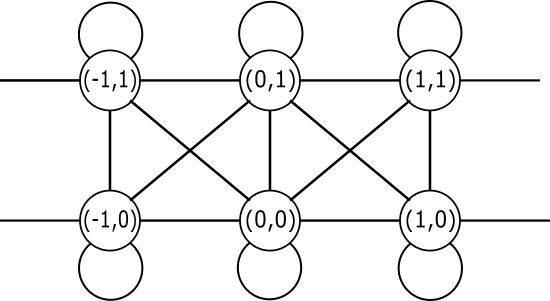}
\centering
\caption{Any quantum walk of a particle on a line with a two dimensional internal degree of freedom can be represented by a quantum walk on this expanded graph. For generality, all links are shown undirected, allowing travel in both directions.}
\label{fig:Internal}
\end{figure}

In particular, suppose that initially the different spatial locations form a directed graph with edge set $E \subseteq \{ n \rightarrow m \, | \, n,m \in V\}$, then we can construct a new graph to represent the situation including the internal degrees of freedom, with vertices $V' = \{ (n,k) \, | \, n \in V, k\in \mathcal{S}_n\}$ and edge set $E' = \{ (n,k) \rightarrow (m, l) \, | \, n \rightarrow m \in E,  k \in S_n, l \in S_m\}$. For example any coined Quantum walk of a particle  on a line  with a two-dimensional degree of freedom is identical to a walk of a particle with no internal degrees  on the  graph shown in figure \ref{fig:Internal}.  

Local probability conservation on the expanded graph then implies local probability conservation for the original graph, with the probabilities and currents on the original graph being $P_n = \sum_k P_{(n,k)}$ and $J_{mn} = \sum_{k,l} J_{(m,l), (n,k)}$

\subsection{Mixed states and general quantum processes} \label{sec:mixed}

So far we have considered pure quantum states evolving unitarily. However, it is also possible to extend these results to mixed states and general quantum processes (represented by completely positive trace preserving maps), which may be useful when considering open quantum systems or situations involving uncertainty. In this case the state is represented by a density operator $\rho$, and the transformation during a single time-step is given by $\rho' = \sum_i K_i \rho K_i^{\dagger}$, where $K_i$ are Kraus operators \cite{nielsen}. In order to respect the locality structure of the graph, such a transformation must satisfy $n \rightarrow m \notin E \implies \bra{m} K_i \ket{n}=0\;\forall\, i$. Mixed states and general quantum dynamics can always be represented by pure states and unitary evolutions on a larger hilbert space composed of the original system and an ancilla \cite{nielsen}. By treating the ancilla as an internal degree of freedom as in the previous subection, it follows that local probability conservation also applies in these cases.

\section{Discussion}

\noindent For quantum evolutions in discrete space and time, in which the locality structure of space is described by an arbitrary directed graph and the evolution is unitary, we have shown that probability is locally conserved. Essentially, we can always explain the change in spatial probability distributions in terms of probability flows which respect the locality  of space. 

The constraint of local probability conservation can  be expressed  in terms of the probability current $J_{nm}$ between vertices or probability flows $f_{nm}$ along edges. Unlike in the continuous time examples which have been considered, the existence of a valid probability flow is established non-constructively, although valid solutions can be obtained efficiently via numerical methods. 

A third approach to the probability flow is to consider a stochastic matrix\footnote{i.e. satisfying $P_{m|n} \geq 0$, and $\sum_m P_{m|n} = 1$ for all $n$}  $P_{m|n}$ which evolves the initial probability distribution into the final distribution via 
\begin{equation} 
P_m' = \sum_n P_{m|n} P_n,
\end{equation} 
with $n \rightarrow m \notin E \; \implies \; P_{m|n}=0$. This is equivalent to the formulation in terms of probability flows. To go from $f_{mn}$ to $P_{m|n}$  we take
\begin{equation} \label{eq:probflow} 
P_{m|n} = \frac{f_{mn}}{P_n}, 
\end{equation} 
whenever $P_n \neq 0$. If $P_n=0$, \eqref{eq:probflow} is not well defined. However, in such cases the distribution $P_{m|n}$ is irrelevant as there is no probability initially at $n$ to flow, and we can simply take $P_{m|n}=\delta_{m,n}$ to avoid violating the locality structure. Similarly we can transform from $P_{m|n}$ to $f_{mn}$ by taking $f_{mn}= P_{m|n} P_n$.

This result could be helpful in understanding quantum walk evolutions, and is also interesting from a foundational perspective, as it demonstrates that an intuitive property of quantum theory in continuous space and time and discrete space continuous time also holds in the discrete space and time formalism. This could be helpful for any approaches to particle physics in which discretization of time and space is pursued, such as \cite{Bial94, Strauch06, farrelly14, dariano14}.

\begin{acknowledgments}

The authors acknowledge helpful discussions with Chris Cade and Ben Schumacher. 

\end{acknowledgments}


\bibliography{maintextbib}

\appendix

\section{Equivalence of flow and current conditions} 

In this appendix, we show that if we can define a $J_{nm}$ which is real, antisymmetric, satisfies \eqref{eqn:DSDT} and \eqref{eqn:Prob_condition}, and for which $J_{mn} > 0$ only if $m \rightarrow n \in E$ then we can always generate flows $f_{mn}$ satisfying conditions \eqref{Flow1}-\eqref{Flow4}. As we showed in the main paper that these flow conditions always allow one to construct a probability current $J_{mn}$ with the specified properties this shows that these two sets of properties are equivalent.

To achieve this,  we set 
\begin{equation} 
f_{mn}= \left\{ \begin{array}{cl} J_{mn} & \mathrm{if} \, J_{mn}>0 \; \mathrm{ and } \; m \neq n  \\ P_n -   \sum_{m^*}J_{m^{*}n} & m=n \\
0 & \mathrm{otherwise}. 
\end{array}  \right.
\end{equation}
Property \eqref{Flow1} is ensured by \eqref{eqn:Prob_condition},  property \eqref{Flow2} follows because $n \rightarrow m \notin E \, \implies \,m\neq n\; \mathrm{and} \; J_{mn} \leq 0 \, \implies \, f_{mn}=0$. The remaining two properties are given by 
\begin{align} 
 \sum_m f_{mn} &=  \sum_{m\neq n} f_{mn} + f_{nn} \nonumber \\ 
 &= \sum_{m^*}J_{m^{*}n} + \left(P_n -   \sum_{m^*}J_{m^{*}n}\right)  \nonumber \\
 &= P_n 
 \end{align} 
 \begin{align} 
 \sum_n f_{mn} &=  \sum_{n\neq m} f_{mn} + f_{mm} \nonumber \\ 
 &= \sum_{n : J_{mn}>0} J_{mn} + \left(P_m -   \sum_{k : J_{km} >0} J_{km}\right)  \nonumber \\
 &= - \sum_{k : J_{km}<0} J_{km} + P_m -   \sum_{k : J_{km} >0} J_{km}  \nonumber \\
 &=- \sum_k J_{km} + P_m \nonumber \\
 &= \Delta P_m + P_m  \nonumber \\
 & = P_m'
 \end{align} 

\end{document}